\documentclass[english,preprintnumbers,amsmath,amssymb,floatfix]{revtex4}
\usepackage[latin9]{inputenc}
\usepackage{graphicx}
\usepackage{amssymb}

\usepackage{babel}
\usepackage{dcolumn}
\usepackage{bm}

\begin{document}

\title{Spin injection and relaxation in a mesoscopic superconductor}

\author{N. Poli$^{1}$, J. P. Morten$^{2}$, M. Urech$^{1}$, Arne Brataas$^{2}$,
D. B. Haviland$^{1}$ and V. Korenivski$^{1}$ }
\affiliation{$^{1}$Nanostructure Physics, Royal Institute of Technology, 10691
Stockholm, Sweden}
\affiliation{$^{2}$Department of Physics, Norwegian University of Science and
Technology, 7491 Trondheim, Norway}

\begin{abstract}
We study spin accumulation and spin relaxation in a superconducting
nanowire. Spins are injected and detected by using a set
of magnetic tunnel contact electrodes, closely spaced along the nanowire. We observe a giant enhancement of the spin accumulation of up to five orders of magnitude on transition into the superconducting state, consistent with the expected changes in the density of states. The spin relaxation length decreases by an order of magnitude from its value in the normal state. These measurements combined with our theoretical model, allow us to distinguish  the individual spin flip mechanisms present in the transport channel. Our conclusion is that magnetic impurities rather than spin-orbit coupling dominate spin-flip scattering in the superconducting state. 
\end{abstract}
\maketitle

Non-local measurement technique \citep{Johnson1985,Johnson1988,Johnson1988b} is a powerful way to directly probe non-equilibrium spin populations.  The technique has been used to uncover a number of spin-dependent phenomena in nanostructures, such as electron spin precession \citep{Jedema:02 416}, spin Hall effect \citep{Valenzuela2006}, and spin injection and propagation in Si \citep{Appelbaum} and graphene \cite{Tombros}.  Experiments reported to date, have focused on normal metals and semiconductors. In this work, we present direct measurements of the spin transport parameters in a superconductor, performed using a multi-electrode nano-device with tunnel junction injection and simultaneous spin-sensitive detection at different mesoscopic distances from the injection point. We observe dramatic changes in the properties of the injected non-equilibrium spins on transition of the nanowire into the superconducting state.  An interpretation of the observed effects is given by extending recently developed theories \citep{Zhao:95,Yamashita:02,Takahashi:03,Morten:04,Morten:05}.

Non-equilibrium superconductivity has been studied since the pioneering experiments on tunnel injection of quasi-particles (QP's) into superconductors (S) from normal (N) and ferromagnetic (F) metals \citep{Clarke:72,Tedrow}. It was found that the injected electrons remain unpaired QP's for about $10\ \mathrm{\mu s}$ before they combine to form Cooper pairs and condense in the superconducting ground state \citep{Levine,Hsieh}. Much faster relaxation occurs of the charge imbalance, the unequal populations of hole like and electron like QP's, resulting from the tunnel injection \citep{Tinkham&Clarke:72,Tinkham:72}. The first experimental study on \emph{spin-dependent} injection and detection in S (Nb) using a non-local measurement configuration indicated a strong reduction in the spin-flip length ($\lambda_{sf}$) at $T<T_{c}$ \citep{Johnson:94}. On the other hand, local measurements on metal stacks containing Nb were used to infer only a small reduction in $\lambda_{sf}$ on transition into the S state \cite{Gu:02}. A two-fold decrease in $\lambda_{sf}$ of Al below $T_{c}$ was estimated by studying injection from F into S and using \textit{spin-independent} detection \citep{Shin:05}.  All of these experiments on spin injection and relaxation in S used metallic contacts between the ferromagnetic electrodes and the superconductor, which is known to lead to proximity effects suppressing the gap in S and strong Andreev processes \citep{Andreev:64}. Furthermore, $\lambda_{sf}$ was not measured directly as the magnitude of spin splitting in S versus the distance from the spin injection point, but rather inferred from the charge transport characteristics. Our device allows us to simultaneously measure the spin splitting at several points along the superconducting nanowire, and thereby directly determine $\lambda_{sf}$ in S, without the complications due to the proximity effects or Andreev processes.

Fig. (1a) shows a scanning electron microscopy (SEM) image of our device, together with a schematic of the measurement arrangement, which is an extension of the configuration first used by Johnson and Silsbee \citep{Johnson1985,Johnson1988,Johnson1988b}. The samples were fabricated using $e$-beam lithography and a two-angle deposition technique. First, a 15-25 nm thick and 100 nm wide Al strip is evaporated at normal incidence and subsequently oxidized in pure oxygen at $80$ mTorr for 15 min to form a thin Al$_{2}$O$_{3}$ layer at the surface. After the oxidation and without breaking the vacuum, three 50-80 nm wide Co electrodes were deposited at an angle to overlap the Al strip.
The overlap is designed in such a way as to prevent the fringing magnetic
fields at the ends of the Co electrodes from penetrating the superconductor.
The Co electrodes have different widths in order to obtain different
switching fields, so one can set all possible parallel and anti-parallel
injector/detector configurations using an external field. Thus, a
set of Co/Al-O/Al tunnel junctions closely spaced along the Al wire
were formed, with a typical tunnel resistance of $\sim$100 $k\Omega$.
Not only does the use of tunnel junctions increase the effective spin
polarization and thereby the spin signal to be detected \citep{Rashba},
it is also important in providing true QP injection and suppressing
Andreev reflection effects.

Due to the difference in the density of states at the Fermi level
in the spin up and spin down sub bands in F, the spin-polarized charge
current injected from the ferromagnetic electrodes into the Al nano-wire
induces a spin accumulation near the injection point. This accumulation decays away
from the injection point due to spin relaxation, as shown in Fig. (1b), with the spatial profile governed by the diffusion equation:

\begin{equation}
\nabla^{2}(\mu_{\uparrow}-\mu_{\downarrow})=\frac{1}{\lambda_{sf}^{2}}(\mu_{\uparrow}-\mu_{\downarrow}),\label{Diffusion Equation}
\end{equation}
where $\mu_{\uparrow(\downarrow)}$ is the electro chemical potentials
for the spin-up and spin-down carrier populations, $\lambda_{sf}=\sqrt{D\tau_{sf}}$
the spin flip length, $D$ the diffusion constant, and $\tau_{sf}$
the spin flip time. The spin accumulation is symmetric about the injection
point \citep{Urech}, so an F electrode placed outside the charge current path at distance $x$ from the injection point will sense a voltage, which is a direct measure of the local spin accumulation. The voltage difference taken between the parallel (P) and anti-parallel (AP) magnetic states of the injector/detector, normalized by the current, defines the non-local spin signal, which in the N-state is given by \citep{Johnson1985,Johnson1988,Johnson1988b,Takahashi:03,Jedema:02 416}:

\begin{equation}
R_{S}(x)=\frac{V_{P}-V_{AP}}{I_{inj}}=P^{2}R_{N}\exp{(-x/\lambda_{sf})},\label{Rs}\end{equation}
 where $P$ is the spin polarization, $R_{N}=\rho\lambda_{sf}/A$
the characteristic resistance of N, $\rho$ the resistivity of Al,
and $A$ the cross sectional area of the Al strip.

A number of novel effects connected with spin injection and relaxation
in superconductors (S) have been predicted recently \citep{Zhao:95,Yamashita:02,Takahashi:03,Morten:04,Morten:05}.
Cooper pairs have zero spin and carry only charge. It is therefore
spin-polarized electrons tunneling into the QP branches that transport
spin in S. Fig. (1a) illustrates the spin accumulation ($\mu_{\uparrow}-\mu_{\downarrow}$) in S due to spin polarized tunneling from F. Observe that the minimum injection energy is the gap energy in the superconductor $\Delta \approx 200 \mu$eV for Al. If the injection energy is close to the gap energy ($\Delta$), then spin-polarized QP's can be created and charge imbalance avoided \citep{Tinkham:72}. A dramatic increase in the spin accumulation
compared to that in the normal state for the same injection current
and near-gap bias is then expected \citep{Takahashi:03}. This is
understood as originating from the reduction in the density of states
of the QP's due to the opening of the gap in the energy spectrum. Considering spin relaxation due to spin-orbit interaction, the proper spin signal in S is obtained by scaling $R_N$ in Eq. \ref{Rs} with the density factor $(2f_{0}(\Delta))^{-1}$,  \citep{Takahashi:03}:

\begin{equation}
R_{S}(x)=P^{2}\frac{R_{N}}{2f_{0}(\Delta)}\exp{(-x/\lambda_{sf})},\label{Super Rs}\end{equation}
 where $f_{0}(E)=1/(\exp(E/kT)+1)$ is the Fermi distribution function
for a given temperature $T$. Thus, a diverging spin signal is expected
in S as $T\rightarrow0$.

More generally, relaxation of the above non-equilibrium spin accumulation in S is governed by two main mechanisms: scattering by spin-orbit interaction and magnetic impurities. In the elastic limit, these two mechanisms have been studied theoretically and are expected to result in a different energy and temperature dependence of the spin flip processes \citep{Morten:04,Morten:05}. Hence, $\lambda_{sf}$ becomes an energy and temperature dependent quantity in the S-state and can not be quantified by a number, but rather by a function. Seemingly a complication, this $\lambda_{sf} (T)$ dependence can be used to distinguish between the different spin relaxation mechanisms in our device, thus leading to a novel spin flip spectroscopy. The specific prediction is that spin flip by magnetic impurities is enhanced for QP energies close to $\Delta$, whereas spin flip due to spin-orbit interaction is the same in S and N states. We assume that the spectral properties of the aluminum are given by the spatially homogeneous BCS solutions with the temperature dependence of the gap $\Delta\approx1.76\: T_{C}\tanh(1.74\:\sqrt{T/T_{C}-1})$, where $t=T/T_{C}$ is the normalized temperature. This assumption is valid when the contacts to the superconductor are of low transparency and of spatial dimensions smaller than the coherence length in S - the geometry chosen in this experiment with $\sim$50 nm scale tunnel contacts. In the linear response limit, the non-local spin signal
at the detector contact at a distance $x$ away from the injection point becomes

\begin{equation}
R_{S}(x)=P^{2}R_{N}\:\frac{g(x/\lambda_{sf},t)}{\chi(t)\: h(t)},\label{Super Rs Morten}\end{equation}
 where $\chi(t)=-2\int_{\Delta}^{\infty}\frac{E}{\sqrt{E^{2}-\Delta^{2}}}\:\frac{\partial f_{0}(E)}{\partial E}\: dE$
is the Yosida function, and $g(x/\lambda_{sf},t)$ and $h(t)$ are
rather complex energy integrals that can be approximated in S as $h(t)\approx(1-P^{2})\:\chi(t)$
and \begin{equation}
g(x/\lambda_{sf})\approx\int\frac{\partial f_{0}(E)}{\partial E}\:\frac{{-4N}^{2}(E)\: e{}^{-x/(\lambda_{sf}\alpha)}}{2\alpha+N(E)R_{N}/R_{I}}\: dE.\label{g}\end{equation}
 Here $R_{I}$ is the injector tunnel resistance, $N(E)$ the density
of states of the superconductor and $\alpha=\sqrt{(E{}^{2}-\Delta^{2})/(E{}^{2}+\beta\Delta{}^{2})}$
gives the renormalization of $\lambda_{sf}$. The parameter $\beta=(\tau_{so}-\tau_{m})/(\tau_{so}+\tau_{m})$,
with $\tau_{so}$ and $\tau_{m}$ being the normal state spin-orbit and magnetic
impurity spin relaxation times, respectively, is a measure of the
relative contributions from the two scattering mechanisms.  $\beta$
is expected to approach 1 if magnetic impurities dominate spin flip
processes, i.e $\tau_{m}\ll\tau_{so}$, which results in a substantial decrease in $\lambda_{sf}$. For dominating spin-orbit induced spin flip, i.e $\tau_{m}\gg\tau_{so}$, $\beta=-1$ which gives $\alpha = 1$, so that there is no renormalization of $\lambda_{sf}$ in Eq. \ref{Super Rs Morten}. The effective $\lambda_{sf}$ can be extracted by fitting the theoretical $R_S$ of Eq. \ref{Super Rs Morten} to the $R_S$ measured by the two detectors placed at 300 and 600 nm.

The multi-electrode nano-device discussed above and illustrated in Fig. 1b is capable of direct measurements of the spin accumulation and the spin flip length, and is therefore ideal for exploring the fundamental properties of spin transported
in S. Measuring the spin signal versus the distance, $x$, from the injection
point, as shown in Fig. 1c, allows a direct determination of the spin flip length,
$\lambda_{sf}$. In our case of multiple spin detectors, this direct measurement of $\lambda_{sf}$ is done in-situ in the same device, in a single field sweep. The measurements were performed using the lock-in technique, with a 7 Hz bias signal applied to the injector and the right end of the Al wire. Typical values of the bias current used were $I_{34}$=5 $\mu$A rms in the N state and $1-10$ nA in the S state of the nano-wire. The non-local voltages $V_{10}$ and $V_{20}$ were
measured using preamplifiers with very high input impedance ($\sim10^{15}~\Omega$)
and low input bias currents ($\sim10$ fA) in order to minimize spurious
contributions to the detected signals. At 4 K the typical junction resistances were 50-200 k$\Omega$ and the thin film Al resistivity 5-10 $\mu\Omega cm$. Using the Einstein relation $\sigma=e^{2}N_{Al}D_{N}$,
with $N_{Al}=2.4\times10^{28}$ eV$^{-1}$m$^{3}$ \citep{Jedema:02 416}
being the density of states at the Fermi level, gives the diffusion
constant $D_{N}=(3-9)\times10^{-3}$ m$^{2}$s$^{-1}$. Fitting the data from typical $R_S$ vs. $H$ (applied magnetic field) curves \cite{Poli,Urech:06} to Eq. \ref{Rs} yields $\lambda_{sf}=800-1100$ nm, $\tau_{sf}\thickapprox100$
ps and the effective spin polarization of $P=12$ \%. These spin transport
parameters in the N state are in good agreement with the recent results
for similar structures \citep{Jedema:02 416,Valenzuela2006,Valenzuela2005,Garzon}.

It is important that the spin channel is maintained superconducting
throughout the magneto-transport measurements. The typical bias current
used for the transport measurements in the S state is $\sim1$ nA, which is low enough not to suppress superconductivity due to QP injection. Moreover, from critical current measurements, similar to those reported previously \cite{Urech:06}, we conclude that  possible changes in the fringing fields have no effect on the superconducting parameters relevant for the spin transport properties discussed below.

Figure 2 shows the normalized $R_{S}$ for sample 1 for
$x=300$ nm as a function of temperature. The bias current was kept
at 1 nA in order not to affect the superconducting gap by the QP injection
\citep{Urech:06}, and to obtain near gap injection energies. $R_{S}$ is enhanced in the S state by 4 to 5 orders of magnitude. This is by far the largest $R_{S}$ measured in a metal/oxide nano-structure. The theoretical fit using Eq. \ref{Super Rs Morten}-\ref{g} approximates well the experimental data for temperatures down to $T\sim0.2T_{C}$, at which point the measured $R_{S}$ starts leveling off. We believe this to be due to an effective QP temperature higher than that given
by the thermometer in the 10-100 mK range \citep{Corlevi}. The QP's are relatively decoupled from the phonon bath at the lowest temperatures. The noise due to the electromagnetic environment in the measurement system, affects the injected QP's and raises their temperature. This heat is not fully dissipated by the phonon bath, since the phonon population vanishes as $T$ approaches zero. In order to determine the effective QP temperature, we model the normalized differential conductance of the injection junction, measured at 22 mK. Using the model of \cite{Giaver1961}, Figure 2 shows the best fit, which was obtained for $T_{eff.}\approx 0.2 T_C$. This value is consistent with the saturation behavior of $R_S$, further supporting our interpretation. Thus, the measured $R_{S}$ saturates as $T \rightarrow 0$, but its dramatic enhancement of 4-5 orders of magnitude is a strong confirmation of the recent theoretical predictions on spin injection in superconductors \citep{Takahashi:03,Morten:05}.

Another key quantity determining spin transport in S is the spin relaxation length, which can be used to differentiate the different spin relaxation mechanisms present in the device. Figure 3 shows the normalized $\lambda_{sf}$ for two samples as a function of temperature. The critical temperature for both samples is $T_{C}\approx1.6$ K and $\lambda_{sf}(T\gtrsim T_{C})\approx1\ \mu$m. The measured $\lambda_{sf}$ decreases substantially at low temperature, by a factor of ten at 20 mK compared to its value in the N state. This temperature dependence of $\lambda_{sf}$ is inconsistent with the behavior predicted for pure spin-orbit scattering \citep{Takahashi:03,Zhao:95} but is in good agreement with the predictions for magnetic impurity mediated spin flip \citep{Morten:04,Morten:05}. The $\lambda_{sf}(T)$ data are well described by the theoretical dependence of Eq.\ref{Super Rs Morten}, as shown in Fig. 4 by the solid line. The best fit was
obtained for $\beta=0.5$, which is equivalent to $1/\tau_{m}=3/\tau_{so}$. This means that spin flip scattering due to magnetic impurities is three times more likely than spin-flip by spin-orbit interaction. With $\beta=0.5$ the renormalization of the scattering rates described by $\alpha$ yields a diverging spin flip rate as $T$ approaches zero, since the spins are injected close to the gap edge, where $\alpha\approx0$.

A magnetic impurity concentration of $\sim1$\% is known to suppress superconductivity \cite{Ronald1973,Aspen1979,Jalkanen2007}, which would manifest in a reduced $T_C$. Our measured $T_C$ is greater than that of pure Al  due to non-magnetic impurity scattering,  typical for thin films (grain boundary and surface scattering).  From this we estimate an upper limit on the concentration of magnetic impurities at 0.1\%. Previous results show that even a magnetic impurity concentration of $0.005$\% can lead to a significant renormalization of $\lambda_{sf}$ in the superconducting state \cite{Aspen1979}.
Thus, the spin flip rate can be significantly enhanced even for low concentrations of magnetic impurities. 

In conclusion, we report direct measurements of the main parameters
of spin transport in a superconductor. The mesoscopic multi-terminal
device used allows an in-situ determination of the spin accumulation
and the spin relaxation length of quasi-particles, which carry the
spin current in the superconducting state. We observe a record high enhancement
of the spin injection efficiency for near-gap bias, up to 4 to 5 orders
in magnitude compared to the normal state, and an order of magnitude
reduction in the spin relaxation length at $T\ll T_{C}$. These effects
are explained theoretically as being due to changes in the quasi-particle
density of states caused by opening of the superconducting gap, and
strong enhancement in spin flip scattering from magnetic impurities at
energies close to the gap energy.

\begin{acknowledgments}
Financial support from the Swedish SSF, Wallenberg and Gustafsson
Foundations are greatfuly acknowledged. 
\end{acknowledgments}

\newpage

\begin{figure}[b]
 \center
\includegraphics[width=0.5\columnwidth]{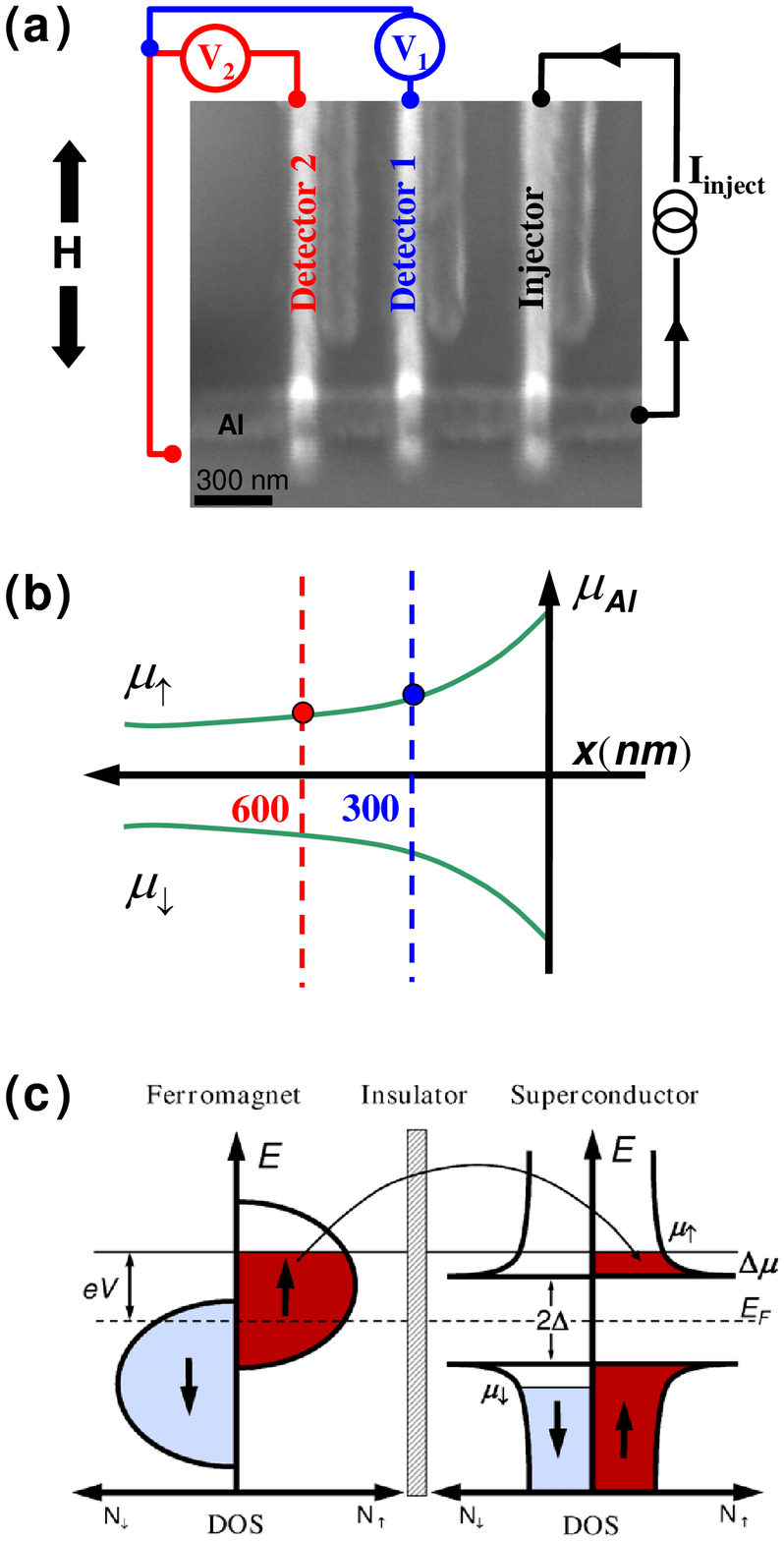}
\caption{ (a) SEM image of the sample together with the non-local measurement configuration. Three vertical Co electrodes are in contact with the horizontal Al strip through tunnel junctions. The electrical circuit schematic illustrates the measurement arrangement used for directly measuring $\lambda_{sf}$. (b) $\lambda_{sf}$ is the exponential decay length of the spin accumulation away from the injection point. (c) Schematic density of states, illustrating spin accumulation ($\Delta \mu=\mu_{\uparrow}-\mu_{\downarrow}$) due to tunneling between a ferromagnet and a superconductor. The largest $\Delta \mu$ is expected for injection energies close to the gap energy, which is  $\approx$200 $\mu$eV for Al. }
\end{figure}

\begin{figure}
\center
\includegraphics[width=0.5\columnwidth]{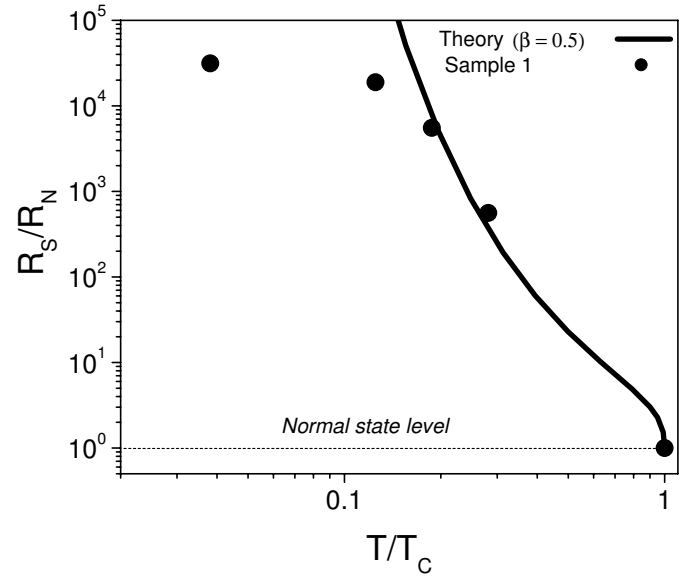}
\caption{Normalized spin signal ($R_{S}$) for sample 1, as a function of normalized temperature. $T_{C}\approx1.6$ K and $R_{S}(4\ \textrm{K})\approx50$ m$\Omega$.}
\end{figure}

\begin{figure}[t]
 \center
\includegraphics[width=0.5\columnwidth]{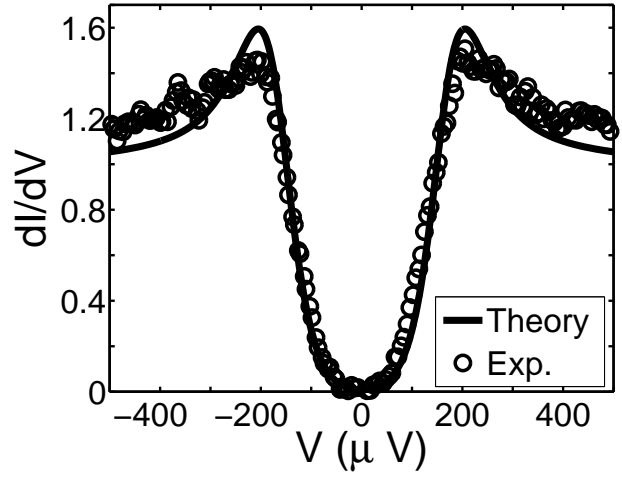}
\caption{Normalized differential conductance of the injection junction measured at 22 mK, together with at theoretical fit \cite{Giaver1961}. The best fit was obtained for $T_{eff} \approx 0.2 T_{C}$. }
\end{figure}

\begin{figure}[t]
 \center
\includegraphics[width=0.5\columnwidth]{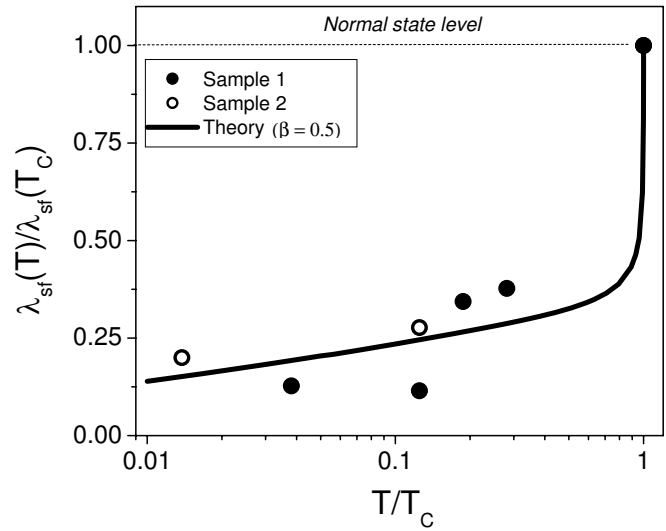}
\caption{\emph{\label{lsf vs T}} Normalized spin diffusion length $(\lambda_{sf})$
for two samples, 20-25 nm in thickness, as a function of normalized
temperature ($T/T_{C}$). $T_{C}\approx1.6$ K and $\lambda_{sf}(4\ \textrm{K})\approx1\ \mu$m. }
\end{figure}

\end{document}